\documentclass{iopconfser}

\usepackage{graphicx}

\begin{document}

\title{Vector meson photoproduction on the nucleus and the extraction of the nuclear suppression factor using asymmetric beams}

\author{Ronan McNulty$^{1}$, Wolfgang Sch\"afer$^{2}$}

\affil{$^1$School of Physics, University College Dublin, Dublin 4, Ireland}

\affil{$^2$Institute of Nuclear Physics, Polish Academy of Sciences,
ul. Radzikowskiego 152, PL-31-342 Krak\'ow, Poland}

\email{ronan.mcnulty@ucd.ie}


\begin{abstract}
    The cross-section for photoproduction of a vector meson on a nucleus
    is usually determined in nucleus-nucleus collisions, which introduces a two-fold ambiguity as to which nucleus emitted the photon.
    This can be resolved in asymmetric collisions of protons and nuclei, where the ratio of photoproduction cross-sections on the proton and nucleus directly measures the nuclear suppression factor with minimal model dependence.
    The feasibilty of using existing and future proton-lead collisions at the LHC is evaluated.
\end{abstract}

\section{Introduction}
\begin{figure}[b]
    \centering
        \includegraphics[scale=0.25]{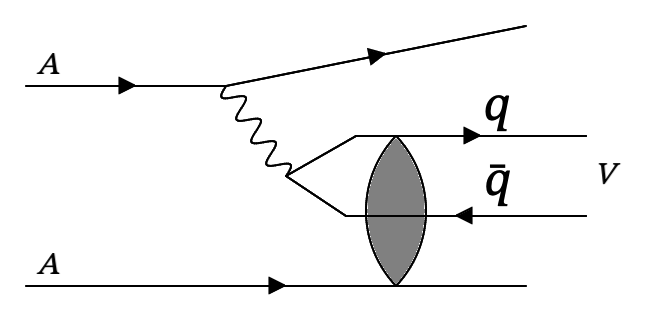}         
    \caption{Graphical representation of photoproduction of a vector meson  in nucleus-nucleus collisions.  The shaded area depicts the interaction of the meson with the target.}
    \label{fig:fd}
\end{figure}

Nuclear shadowing and saturation effects can be probed in 
photoproduction of vector mesons on nuclei~\cite{Brodsky:1994kf,  Klein:2019qfb, Guzey:2020ntc, Accardi:2012qut}.
The process is usually studied in ultraperipheral collisions (UPC)~\cite{Bertulani:2005ru} of pairs of nuclei when a photon is emitted from one nucleus, fluctuates into a virtual quark pair, and collides with the other target nucleus, as shown in Fig.~\ref{fig:fd}.
The experimentally measured cross-section, $\sigma_{{\rm AA}\rightarrow {\rm A}\,V\, {\rm A}}$, combined with a calculation of the photon flux, gives the photoproduction cross-section on the nucleus, $\sigma_{\gamma {\rm A}\rightarrow V {\rm A}}$.
Comparing this with experimental results on photoproduction on the proton,
$\sigma_{\gamma p\rightarrow V p}$, allows the nuclear suppression factor to be determined~\cite{Guzey:2020ntc, Mantysaari:2023xcu}.

In the impulse approximation~\cite{Chew:1952fca}, the photon couples coherently to the nucleus but each nucleon is treated as an independent scattering centre, so the ratio 
 $\sigma_{\gamma {\rm A}\rightarrow V {\rm A}}$ to 
$\sigma_{\gamma p\rightarrow V p}$
scales with the number of nucleons to the power of $\frac{4}{3}$.
However, at high energies the photon transforms into a virtual
meson before it hits the target~\cite{Ioffe:1969kf} and multiple interactions can occur between the meson and the nucleons leading to a suppression of  $\sigma_{\gamma {\rm A}\rightarrow V {\rm A}}$ compared to the impulse approximation.
These multiple interactions can be described by the Gribov--Glauber mechanism~\cite{Gribov:1968jf,Glauber:1970jm} giving predictions~\cite{Guzey:2016piu} where  the suppression depends on the total cross-section for the meson to interact with the nucleon.
An observation of suppression in excess of these predictions would point to additional mechanisms for
nuclear shadowing~\cite{Guzey:2016qwo}, most notably saturation, where the gluon densities become so large that non-linear QCD effects, due to gluon recombination, become important.

The production of a vector meson of mass $m_V$ at rapidity $y$ at a centre-of-mass energy of the nucleon-nucleon system $\sqrt{s_{NN}}$ probes values of the energy scale $Q^2=m_V^2$ and fractional momentum of the partons in the nucleon $x=m_T e^{-y}/\sqrt{s_{NN}}$, where $m_T = \sqrt{p_T^2 + m_V^2}$), with $p_T$ the momentum of the meson and in nuclear collisions, $p_T<<m_V$.

Indicative values for the onset of saturation are often given by the saturation scale~\cite{Golec-Biernat:1998zce}  
\begin{equation}
 Q_S^2={Q_0^2}\biggl(\frac{x_0}{x}\biggr)^\lambda   
\end{equation}
with fits to deep inelastic data giving $x_0\sim10^{-4}$ and $\lambda\sim0.3$ at the reference value $Q_0=1$ GeV.  
Thus, for the $J/\psi$ meson,  saturation would occur for $x$ values below about $10^{-7}$, while for the $\rho$ meson it could be seen below $5\times10^{-4}$. 
However, saturation is expected to be enhanced in nuclear collisions by a factor $N_A^{1/3}$ due to the increased gluon density~\cite{McLerran:1993ni,Accardi:2012qut}, where $N_A$ is the number of nucleons in the ion.

Measurements of the photoproduction cross-section in UPC nucleus-nucleus collisions have been performed for the $\rho^0$-meson by the STAR collaboration~\cite{STAR:2007elq, STAR:2011wtm, STAR:2017enh}
at  $\sqrt{s_{\rm NN}}$, of 62.4 and 200 GeV on gold nuclei, 
by the ALICE collaboration on xenon at 5.44 TeV~\cite{ALICE:2021jnv} and lead at 2.76 TeV, while photoproduction on lead at 5.02 TeV has been measured by both the ALICE ~\cite{ALICE:2015nbw,ALICE:2020ugp,ALICE:2023jgu}
and LHCb~\cite{LHCb:2025fzk} collaborations.
Photoproduction of the $J/\psi$ meson has been performed by the STAR collaboration~\cite{STAR:2023vvb, STAR:2023nos} and at the LHC by the ALICE~\cite{ALICE:2021tyx,ALICE:2021gpt,ALICE:2018oyo}, ATLAS~\cite{ATLAS:2025aav}, CMS~\cite{CMS:2023snh} and LHCb~\cite{LHCb:2022ahs} collaborations.
Both the $\rho$ and $J/\psi$ meson measurements show nuclear suppression factors somewhat in excess  of that expected by the Gribov-Glauber mechansims.  The $J/\psi$ measurements are of particular interest since perturbative calculations can be performed in QCD with and without models that include saturation.  The data show suppression that may even exceed that expected by saturation~\cite{Guzey:2013xba,Mantysaari:2023xcu}, although suppression can also be obtained by the inclusion of higher Fock states~\cite{Luszczak:2024kgi}.

The search for saturation is one of the main motivations for the electron-ion collider (EIC) currently under construction~\cite{Accardi:2012qut}.
Photoproduction at the EIC has an advantage over photoproduction in nucleus-nucleus collisions since the photon is emitted from the electron, while there is a two-fold ambiguity in nucleus-nucleus collisions.
On the other hand, the centre-of-mass energy of the EIC limits the reach to $x$ values of about $10^{-4}$, while in lead-lead collisions at the LHC $x$ values down to $10^{-6}$ are probed.

The two-fold ambiguity does not occur at a rapidity of zero, providing one $x$ value at which the nuclear suppression can be determined.  However, this does not probe the lowest $x$ values, which are only accessible in the forward region and are probed by the higher energy photon. 
The two-fold ambiguity can be lifted by tagging the emission of forward neutrons~\cite{Rebyakova:2011vf,Guzey:2013jaa} that occur due to simultaneous photon exchanges that lead to electromagnetic excitations of the nucleus~\cite{Baltz:2002pp}.
Neutron emission is therefore more probable at low impact parameters and thus preferentially accompanies the higher energy photon.
This technique has been implemented by STAR~\cite{STAR:2023nos}, ALICE~\cite{ALICE:2018oyo} and CMS~\cite{CMS:2023snh} making use of their zero-degree calorimeters.
However, theoretical uncertainties remain in the  modelling of the neutron emissions~\cite{Jucha:2025wet,Dyndal:2026uvm}.

In this paper, an alternative approach is suggested that makes use of asymmetric proton-nucleus collisions.
In  2016 during Run 2 of the LHC, protons and lead ions were collided and about 30~nb$^{-1}$ of data were collected by LHCb, 40~nb$^{-1}$ by ALICE, and 200~nb$^{-1}$ by ATLAS and CMS~\cite{Jowett:2017dqj}.  
In asymmetric collisions, the lead ion  is nearly always the photon emitter as the photon flux is enhanced by the square of the charge on the nucleus.  Thus photoproduction on the proton is the dominant production mechanism.
However, photoproduction on the nucleus also occurs with a total cross-section that is about 5\% that on the proton.  
Critically, the kinematic distributions of these events is different allowing the two classes to be disentangled.
Therefore the asymmetric data already taken has the potential to simultaneously measure photoproduction on the proton and on the nucleus.

In section~\ref{sec:th} the relative cross-sections for photoproduction on the ion and on the proton are calculated and two observables are presented: one that minimises experimental uncertainties and one that minimises theoretical uncertainties.
Section~\ref{sec:mc} then presents a study using pseudo-data produced with the STARlight generator~\cite{Klein:2016yzr} showing that these measurements can be performed with existing data and the  nuclear suppresion factor can be directly measured.
Section~\ref{sec:conclude} presents conclusions.

\section{Predictions for photoproduction on the ion compared to on the proton}
\label{sec:th}

The differential cross-section for ${\rm AA}\rightarrow {\rm A}+V+ {\rm A}$ where $A$ is an ion and $V$ is a vector meson can be expressed in terms of the photoproduction cross-section $\sigma_{\gamma {\rm A}\rightarrow V{\rm A}}$ using the photon flux $dN_\gamma/dk$:
\begin{equation}
\label{eq:pbpbgpb}
{d\sigma_{{\rm AA}\rightarrow {\rm A} V {\rm A}}\over dy}
=
\biggl(k^+{dN_\gamma\over dk^+}\biggr)\sigma_{\gamma A\rightarrow V A}(W^+)
+
\biggl(k^-{dN_\gamma\over dk^-}\biggr)\sigma_{\gamma A\rightarrow V A}(W^-).
\label{eq:dsdy}
\end{equation}
The two terms on the RHS.\ correspond to each of the ions being the photon emitter.
The photon energy $k^{\pm}=(m_V/2)e^{\pm y}$, where $m_V$ is the mass of the meson and $y$ is the rapidity in the centre-of-mass frame.  The centre-of-mass energy of the photon-nucleon system, $W=(2k\sqrt{s_{NN}})^{0.5}$, where $\sqrt{s_{NN}}$ is the centre-of-mass energy of the nucleon-nucleon system.
By calculating $\sigma_{\gamma {\rm A}\rightarrow V{\rm A}}$ and comparing it to a measurement of $\sigma_{\gamma {\rm p}\rightarrow V{\rm p}}$,  the nuclear suppression factor can be extracted.
However, there is a two-fold ambiguity in Eq.~\ref{eq:dsdy} as to which ion is the photon emitter and which is the target.  
Therefore, in this paper we consider asymmetric proton-nucleus collisions where the differential cross-section is written
\begin{equation}
{d\sigma_{{\rm pA}\rightarrow {\rm p} V {\rm A}}\over dy}
=
\biggl(k^A{dN_\gamma\over dk^A}\biggr)^A\sigma_{\gamma p\rightarrow V p}(W^A)
+
\biggl(k^p{dN_\gamma\over dk^p}\biggr)^p\sigma_{\gamma A\rightarrow V A}(W^{p}).
\label{eq:ppb}
\end{equation}
The first term on the right-hand side corresponds to the photon coming from the nucleus, while the second term corresponds to the photon coming from the proton.
Generally the first term dominates due to the $Z^2$ enhancement of the photon flux from an ion with $Z$ protons.
The rapidity is defined in the centre-of-mass frame, measured with respect to the proton direction, so $k^p=(m_V/2)e^{y}$ and $k^A=(m_V/2)e^{-y}$.
Note that in asymmetric collisions, $y$ is shifted in the proton direction by $\Delta=\frac{1}{2}\ln\biggl(E_p/(E_A/N_A)\biggr)$ where $E_p,E_A$ are the energies of the proton and ion beams, respectively.
Thus, the rapidity
as measured in the laboratory frame
\begin{equation}
    y_{LAB}=y+\Delta.
\end{equation}

The photon number density per unit energy and per unit transverse area can be calculated in the equivalent photon approximation~\cite{Bertulani:1987tz} as a function of photon energy $k$ and impact parameter $b$:
\begin{equation}
    \frac{d^3N}{d^2bdk}=\frac{Z^2\alpha}{\pi^2}
    \frac{k}{\gamma^2}\biggl(K_1^2(\xi)+\frac{1}{\gamma^2}K_0^2(\xi)\biggr),
    \label{eq:flux3}
\end{equation}
with $\xi=bk/\gamma$, the boost is $\gamma=\sqrt{s_{NN}}/(2m_p)$ with $m_p$ the proton mass, and $K_0,K_1$ are modified Bessel functions.
Restricting $b$ to be greater than the radius of the proton, $R_p$, plus the radius of the ion, $R_A$, as is required in ultraperipheral collisions, gives
\begin{equation}
    k{dN_\gamma(k)\over dk}
    =
    {2Z^2\alpha\over\pi}
    \biggl(\xi K_0(\xi)K_1(\xi)-{\xi^2\over2}[K_1^2(\xi)-K^2_0(\xi)]\biggr)\biggl|_{\xi=(R_A+R_p)k/\gamma}.
    \label{eq:flux}
\end{equation}
Therefore for photons of energy, $k$, the fluxes scale as
\mbox{$\biggl(k{dN_\gamma\over dk}\biggr)^{A}= Z^2\biggl(k{dN_\gamma\over dk}\biggr)^{p}$}.

Photoproduction on the nucleon, in the absence of nuclear effects, is given in the impulse approximation by 
\begin{equation}
\sigma^{IA}_{\gamma A\rightarrow V A}(W)
=\frac{d\sigma_{\gamma p\rightarrow Vp}}{dt}(W,t=0)\Phi_A(t_{min})
\label{eq:ia}
\end{equation}
where $\Phi_A$ is the integral over the nuclear form-factor, $F_A$ squared,
\begin{equation}
    \Phi_A(t_{\rm min})=\int_{t_{\rm min}}^\infty |F_A(t)|^2 dt.
    \label{eq:Phi}
\end{equation}
with $t_{\rm min}=m_V^4m_p^2/W^4$.
The form-factor is a Fourier transform of the nuclear density
\begin{equation}
F_A(q)=\frac{4\pi}{q}\int_0^\infty r\sin(qr)\rho(r) dr
\end{equation}
with $\rho(r)=\rho_p(r)+\rho_n(r)$ where $\rho_p$ and $\rho_n$ are the density of protons and neutrons in the ion and are often described using a Woods-Saxon distribution
\begin{equation}
\rho_N(r)=\frac{\rho_0}{1+\exp((r-R)/d)}
\end{equation}
with $R$ being the radius, $d$ the skin depth, and $\rho_0$ a normalisation constant so that
$\int\rho_pd^3r=Z$ and
$\int\rho_nd^3r=N_A-Z$.
The $t$-differential vector meson cross-sections on the proton have been measured at HERA~\cite{H1:2020lzc,H1:2013okq} and the LHC and to a good approximation can be described by an exponential function
\begin{equation}
\label{eq:pt}
\frac{d\sigma_{\gamma p\rightarrow Vp}}{dt}=\sigma_{\gamma p\rightarrow Vp}b_Ve^{b_Vt}    
\end{equation}
where the slope parameter, $b_V$ depends on $W$, and at
HERA energies, $b_{J/\psi}\sim 4$GeV$^{-2}$, $b_{\rho}\sim 10$GeV$^{-2}$.    
Eq.~\ref{eq:ppb} can therefore be written
\begin{equation}
{d\sigma_{{\rm pA}\rightarrow {\rm p} V {\rm A}}\over dy}
=
Z^2\biggl(k_A{dN_\gamma\over dk_A}\biggr)^p\sigma_{\gamma p\rightarrow Vp}(W_A)
+
\biggl(k_p{dN_\gamma\over dk_p}\biggr)^p
b_V\sigma_{\gamma p\rightarrow Vp}(W_p) 
{\cal S}^2
\Phi_A(t_{min})
\label{eq:me}
\end{equation}
where ${\cal S}^2$ is the amount by which the impulse approximation must be scaled and $\cal{S}$ is conventionally defined as the nuclear suppression factor.~\footnote{The rationale for using the term ${\cal S}^2$ to describe the suppression of the vector meson comes from previous work in the literature where, in the perturbative regime of the $J/\psi$ meson, at leading order the cross-section scales with the gluon density squared.  Thus, $S$ can then be interpreted as a gluon suppression factor.}
In $J/\psi$ photoproduction on lead and gold nuclei ${\cal S}^2$ has been observed to vary between 0.25 and 1 with energy~\cite{Mantysaari:2023xcu}, which could be evidence for saturation.
In $\rho$ production, because of the enhanced cross-section for the total $\rho$-nucleon cross-section, re-scattering effects give greater suppression, and ${\cal S}^2\sim0.1$~\cite{Frankfurt:2015cwa}.
The photon fluxes in Eq.~\ref{eq:me} are the number of photons coming from a single proton at a distance greater than $R_p+R_{A}$.

The relative number of events in which the photon comes from the proton
compared to when it comes from the ion, as observed in the laboratory frame, is 
\begin{equation}
\label{eq:rat1}
F_{p/A}(y_{LAB})\equiv
\frac
{N_{\gamma-from-p}(y)}
{N_{\gamma-from-A}(y)}=
\frac
{b_V{\cal S}^2\Phi_A(t_{min})}
{Z^2}
\frac
{\sigma_{\gamma p\rightarrow Vp}(W_p)}
{\sigma_{\gamma p\rightarrow Vp}(W_A)}
\frac
{\biggl(k_A{dN_\gamma\over dk_A}\biggr)^p}
{\biggl(k_p{dN_\gamma\over dk_p}\biggr)^p}
\label{eq:frac}
\end{equation}
with $k_p,k_A,W_p,W_A$ calculated as functions of $y$.
This quantity minimises the experimental uncertainties because
the yields are determined in the same portion of the detector and so acceptance and selection efficiencies largely cancel in the ratio.

Predictions of $F_{p/A}$ for lead ions, along with their uncertainties are calculated  for
$\rho$ and $J/\psi$ mesons
using measured values and uncertainties for the cross-section and $b_V$ parameter and a theoretical calculation of the photon flux.

For the $J/\psi$ meson, the cross-section, measured at HERA~\cite{H1:2013okq}, is well described by a power-law fit of the form
\begin{equation}
    \sigma\sim W^\delta,
\end{equation} with 
$\delta=0.67\pm0.03$.  This also gives good agreement to data at the LHC and is valid over an energy range from $W$ of 30 to 1000 GeV~\cite{LHCb:2024pcz}.
For the $\rho$ meson, a fit to fixed target and HERA data in the $W$ range from 2 to 200 GeV can be fitted with the sum of two power laws, one to describe Pomeron exchange and the other for Reggeon exchange,
\begin{equation}
    \sigma\sim\biggl(\frac{W}{W_0}\biggr)^{\delta_{\cal P}}+f_{\cal R}\biggl(\frac{W}{W_0}\biggr)^{\delta_{\cal R}}    
\end{equation}
with the reference scale $W_0$ fixed to 40 GeV, and $\delta_{\cal P}=0.207\pm0.015$,
$\delta_{\cal R}=-1.45\pm0.12$
and $f_{\cal R}=0.020\pm0.007$~\cite{H1:2020lzc}.

The slope parameter, $b_V$, also depends on energy, which according to Regge theory, can be parametrised as 
\begin{equation}
    b_V=b_0+4\alpha^\prime\log(W/W_0),
\end{equation}
where $\alpha^\prime$ is the slope of the Pomeron trajectory.
Fits to the $J/\psi$ data from HERA and the LHC give $b_0=4.80\pm0.25$\ GeV$^{-2}$ and $\alpha^\prime=0.133\pm0.025$\ GeV$^{-2}$ at the reference scale $W_0=90$ GeV~\cite{LHCb:2024pcz}.
Fits to the $\rho$ data at HERA give 
$b_0=9.59\pm0.14$\ GeV$^{-2}$ and $\alpha^\prime=0.233\pm0.064$\ GeV$^{-2}$ where the reference scale is taken at $W_0=40$ GeV~\cite{H1:2020lzc}.

The photon flux is known to a few percent at all but the highest energies and an uncertainty is estimated by changing the ion radius by 0.5 fm.

\begin{figure}
      \centering
        \includegraphics[scale=0.7]{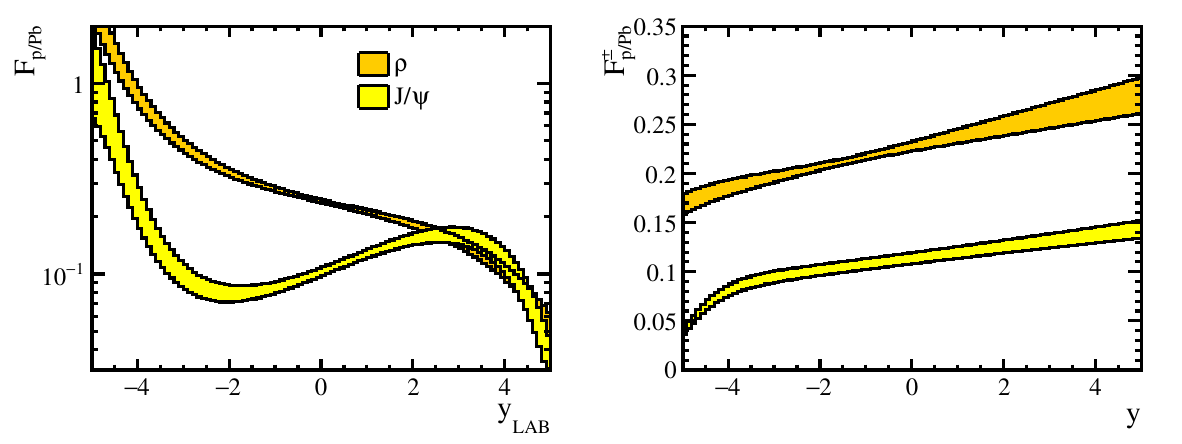}
    \caption{Left: Fraction of events due to the photon coming from the proton compared to lead in proton-lead collisions, assuming no nuclear suppression.
    Right: Number of events with the photon coming from the proton divided by the number of events with the photon coming from the lead ion at negative rapidity.
    \label{fig:frac}}
\end{figure}

The fraction of events where the photon comes from the proton compared to when it comes from the lead ion is shown in the left panel of Fig.~\ref{fig:frac}, where the band indicates $\pm1\sigma$ uncertainties.
Assuming no nuclear suppression (${\cal S}=1$), 
the fraction of $J/\psi$ events where the photon comes from the proton rather than the lead ion
is about 10\% for $|y_{LAB}|<4$ while for the $\rho$  meson it is higher, particularly in the backwards direction due to the enhancement of the cross-section in the Reggeon region at low $W$. 
With ${\cal S}^2\sim0.5$ for the $J/\psi$ and $\sim0.1$ for the $\rho$ meson, this means a few percent of events will have a photon from the proton across most of the rapidity range.
A few-percent effect would be difficult to see when measuring the overall cross-section measurement, especially given the uncertainties in the $\gamma p$  cross-sections.
However, the two categories of events can be distinguished experimentally due to their very different transverse momentum distributions.
In a coherent interaction on a lead target the interactions occurs with the whole nucleus and thus $p_T\sim\hbar c/R\sim40{\rm\ MeV}$, while when the photon comes from the lead ion the size of the proton target determines the typical momentum to be $\sim250 {\rm\ MeV}$.
A measurement of the relative proportions of exclusive events with very low transverse momentum is thus a measurement of Eq.~\ref{eq:frac}, which depends on the nuclear geometry, the photon flux, 
the $\gamma p$ total and $t$-differential cross-sections and the nuclear suppression factor.

Most of these theoretical uncertainties however, can be removed using the data itself by considering the ratio, $F_{p/A}^\pm$, of the 
number of $\gamma$-from-p events at rapidity $y$, 
to the number of $\gamma$-from-A events at rapidity $-y$:

\begin{equation}
F_{p/A}^\pm(y)\equiv
\frac
{N_{\gamma-from-p}(y)}
{N_{\gamma-from-A}(-y)}=
\frac
{b_V{\cal S}^2\Phi_A(t_{min})}
{Z^2}
\label{eq:rat2}
\end{equation}
which directly measures the nuclear suppression factor with minimal model dependence.
This is plotted in the right panel of Fig.~\ref{fig:frac} where the dependence with rapidity is mostly determined by the energy dependence of $b_V$ i.e. the shrinkage of the diffraction peak.

Although theoretical uncertainties mostly cancel in the ratio described in Eq.~\ref{eq:rat2}, experimentally the measurement is slightly more challenging than the ratio of Eq.~\ref{eq:rat1}.
The requirement in Eq.~\ref{eq:rat2} to correlate events at forward and backwards rapidities was neatly addressed by the running condition of the LHC that collided protons and ions in both directions.
Thus an asymmetric detector, such as LHCb, could sample both forward and backward rapidities.  
However, since the laboratory frame does not coincide with the centre-of-mass frame, the theoretically symmetric events in Eq.~\ref{eq:rat2} enter slightly different regions of the detector and thus will have different acceptance effects and selection efficiencies that need to determined using a full detector simulation.

\section{Simulation and expected experimental precision}
\label{sec:mc}
The feasibility of measuring the observables $F_{p/A}$ and $F^\pm_{p/A}$ was evaluated using the {\tt STARlight}~\cite{Klein:2016yzr} MonteCarlo simulation, which was configured to replicate proton-lead collisions at the LHC at $\sqrt{s_{NN}}$=8.16 TeV, with a proton beam energy of 6.5 TeV and a lead beam energy of 2.51 TeV per nucleon.
For the photon coming from the nucleus, {\tt STARlight} uses a numerical integration of Eq.\ref{eq:flux3} over $b$, which includes the probability for having no hadronic interaction at a given value of impact parameter.
For the photon coming from the proton, it uses Eq.~\ref{eq:flux}.
Samples of $\rho\rightarrow\pi\pi$ and $J/\psi\rightarrow\mu\mu$ were generated with an integrated luminosity of 200 nb$^{-1}$, corresponding to what was already delivered to ATLAS and CMS in Run 2 of the LHC. 
For each parent meson, events were simulated with the photon coming either from the lead nucleus or the proton, and combined according to the respective cross-sections.  
The cross-section for photoproduction on the nucleus was generated in two different configurations: firstly, according to the impulse approximation (Eq.~\ref{eq:ia}) and thus with no nuclear suppression; and secondly, with the default option in the generator, 
\begin{equation}
\sigma^{STARlight}_{\gamma A\rightarrow V A}(W)
=\frac{d\sigma_{\gamma A\rightarrow VA}}{dt}\biggl|_{t=0}\int_{t_{\rm min}}^\infty |F_A(t)|^2 dt,
   \end {equation}
where $\frac{d\sigma_{\gamma A\rightarrow VA}}{dt}\biggl|_{t=0}$ is determined using vector meson dominance, the optical theorem, and a Glauber calculation~\cite{Klein:1999qj}.

For the fiducial acceptance of the LHC detectors with full tracking and particle identification in the central region, it was required that the parent meson was in the range $-2<y_{LAB}<2$ and both daughter particles had pseudorapidities, $\eta$, in the range $|\eta|<2$.
For the forward region, it was required that 
$2<y_{LAB}<5$ for the parent and $2<\eta<5$ for the daughters.
Note that during the LHC asymmetric running, both proton-lead and lead-proton orientations were provided meaning that the LHCb detector, which is only instrumented between 
$2<y_{LAB}<5$ could access both
$1.535<y<4.535$ and $-5.465<y<-2.465$.
Within the acceptance a detection efficiency of 70\% was assumed to take account of trigger, reconstruction and selection efficiencies.

\begin{figure}
      \centering
        \includegraphics[scale=0.7]{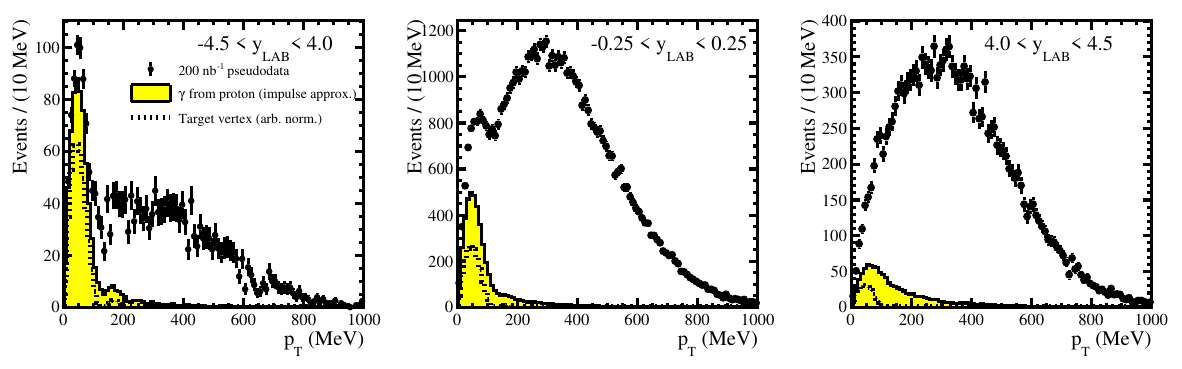}
    \caption{Number of expected reconstructed $J/\psi$ mesons in 
    200 nb$^{-1}$ of proton-lead collisions at 8.16 TeV, in three rapidity ranges.
    The pseudodata has been generated with STARlight and includes an estimate of acceptance and efficiency requirements on the final-state particles, as described in the text. The fraction of events where the photon comes from the proton is indicated by the shaded histogram.  For this contribution, the shape due to the $p_T$ at the target vertex is shown by the dotted line, with an arbitrary normalisation.
    \label{fig:pt}}
\end{figure}

The number of events expected as a function of $p_T$ for $J/\psi$ mesons is shown in Fig.~\ref{fig:pt} for three rapidity regions.
The majority of each distribution is due to events where the photon comes from the nucleus and is consistent with the shape described by Eq.~\ref{eq:pt}.
However, particularly at negative rapidities, a prominent peak can be observed below 100 MeV, which is distinctive of events where the photon comes from the proton.
The shaded histogram shows the contribution from these events alone.  
The $p_T$ is the vector sum of the transverse
momenta at the photon-emission and target vertices
and is principally determined by the latter where the amplitude as a function of $p_T$ is a  Fourier transform of the amplitude in $b$-space, which depends on the shape of the nucleus~\cite{Khoze:2019xke}, and is
shown by the dotted line.
The contribution from the $p_T$ at the photon-emission vertex depends on the proton form factor and is only significant at the highest photon energies.

\begin{figure}
      \centering
        \includegraphics[scale=0.7]{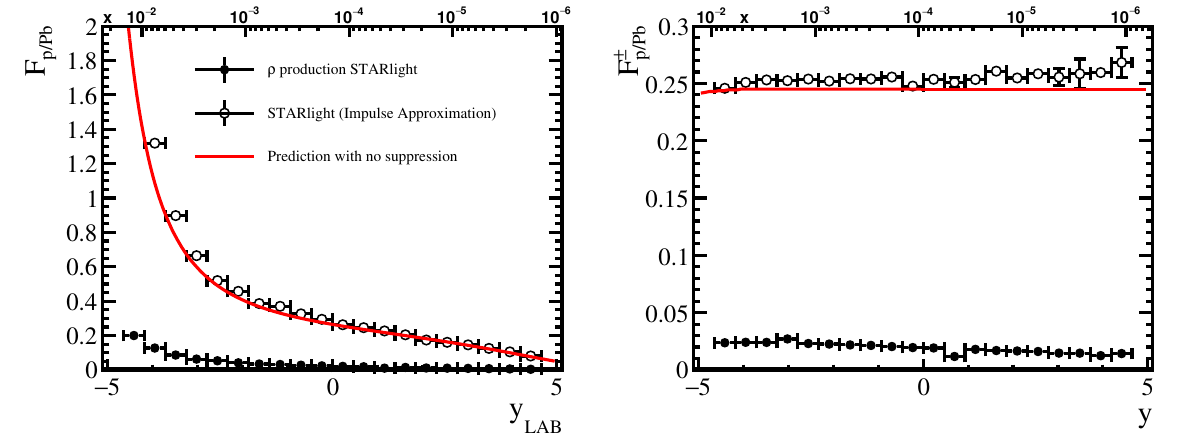}                \includegraphics[scale=0.7]{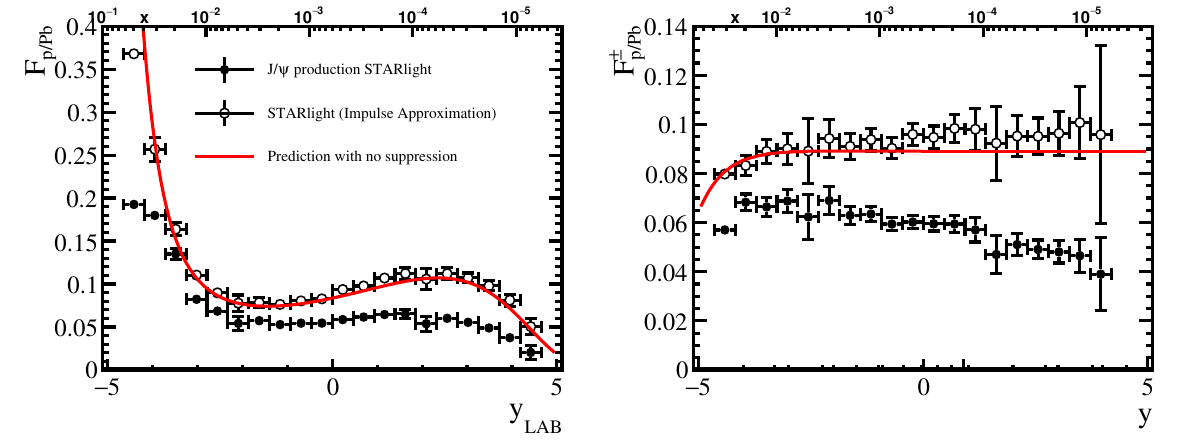}

    \caption{Fitted values for $F_{p/A}$ (left) and $F_{p/A}^\pm$ (right) assuming 200 nb$^{-1}$ of data generated with STARlight using the default configuration and with the Impulse Approximation.  
    The upper panels are for the $\rho$ meson and lower panels are for the $J/\psi$ meson.
    The curves are the predictions of Eq.~\ref{eq:rat1} and \ref{eq:rat2}.
    The horizontal axis shows rapidity on the bottom and Bjorken-$x$ on the top.
    \label{fig:sim}}
\end{figure}

The pseudodata generated by STARlight was analysed as follows.
First, a fit was performed to the $p_T$ distributions, in bins of rapidity, using template shapes for 
the two contributions.
This gives the estimated number of events in the sample where the photon came from the proton, and where the photon came from the ion.  
Second, these numbers were corrected for the detector acceptance and efficiency to give the estimated number of events produced at a given rapidity.
Finally, these yields were combined to determine $F_{p/A}(y_{LAB})$ and $F^\pm_{p/A}(y)$.
The results are shown in Fig.~\ref{fig:sim}.
Superimposed on the data are the expectations from
Eq.~\ref{eq:rat1} and \ref{eq:rat2}, assuming no suppression.
When calculating 
Eq.~\ref{eq:rat1} and \ref{eq:rat2}
the cross-section was assumed to have the energy dependence used in the STARlight generator and $b_V$ was taken to be constant, as there is no energy-dependence implemented in the generator.
Thus the theoretical shapes are slightly different from the central values shown in Fig.~\ref{fig:frac}. 

The data generated with the impulse approximation agree with the theory curves, thus validating the approach.
The size of the error bars on the data determination indicates the likely precision that can be obtained in measuring these ratios with 200 nb$^{-1}$ of data.  
The data that was generated with the default STARlight configuration (i.e. with nuclear suppression due to the Glauber mechanism implemented in the generator) show much lower values for the ratios than that generated in the impulse approximation.  
Therefore, a measurement of these quantities can measure the amount of nuclear suppression as a function of energy.  
With 200\ nb$^{-1}$ of data, a 
statistical precision of about 10\% looks to be feasible for the $J/\psi$ meson, which, since the $J/\psi$ is in the perturbative regime, can be compared to QCD predictions with and without saturation.
For the $\rho$ meson, a 1\% measurement is, in principle, achievable due to the large cross-section. 
In practice, knowledge of the detector acceptance and selection efficiency in the different regions measured in the forward and backward direction, probably limit the precision to a few percent.
However, the $\rho$ is not in the perturbative regime, so it is more difficult to make quantitative comparisons with saturation models.

Of particular importance in the search for saturation, is the amount of suppression at high photon energy (low $x$).  
As can be seen from the upper horizontal axis in Fig.~\ref{fig:sim}, 
here the forward region is of particular importance.  
In Run 2 at the LHC, the luminosity delivered to LHCb was only 13 nb$^{-1}$ at positive rapidities and 19 nb$^{-1}$ at negative rapidities, which is probably insufficient to measure $F^\pm$ precisely. 
Thus, it is highly desirable that another proton-lead run be performed at the LHC~\cite{dEnterria:2025jgm}, where projections for Run 3 and Run 4 foresee 1.2 pb$^{-1}$ of data at ATLAS
and CMS, and 0.6 pb$^{-1}$ at ALICE and LHCb~\cite{Citron:2018lsq}.

\section{Conclusions}
\label{sec:conclude}
In proton-ion collisions at the LHC,
most photoproduced vector mesons are
due to the photon coming from the ion and thus measure
$\sigma_{\gamma p\rightarrow Vp}$
at energy scales from about 10 to 1000 GeV, corresponding to rapidities from -5 to 5.
However, an excess of events at low $p_T$ are due to the photon coming from the proton, which allows a measurement of
 $\sigma_{\gamma A\rightarrow VA}$.
The ratio of these two cross-sections 
allows a direct determination of the nuclear suppression factor.
In the forward region, $x$ values down to $10^{-6}$ are probed, where saturation effects are expected to be important, particularly in the gluon-rich environment of nuclei.
Fitting 200 nb$^{-1}$ of pseudodata generated with the STARlight simulation, it is estimated that a precision of 10\% is obtainable on the nuclear suppression factor using $J/\psi$ mesons, while a precision of 1\% is possible using $\rho$ mesons.
More data, particularly in the forward region, is desirable in order to make precise measurements of nuclear suppression at the lowest $x$-values.

\section*{Acknowledgements}
The work of W.S. was supported by the Polish National Science Center\\
Grant No. UMO2023/49/B/ST2/03665.

\bibliographystyle{unsrt}
\bibliography{main}

\end{document}